\documentclass[conference]{IEEEtran}
\IEEEoverridecommandlockouts
\usepackage{cite}
\usepackage[numbers,sort&compress]{natbib}

\usepackage[utf8]{inputenc}
\usepackage{amsmath,amssymb}
\usepackage{graphicx}
\usepackage{bm}
\usepackage{enumerate}
\usepackage{comment}
\usepackage{xcolor}
\usepackage{url}
\usepackage{color,soul}
\usepackage{subcaption}
\usepackage{float}

\definecolor{light-gray}{gray}{0.8}

\usepackage{amsmath,amssymb,amsfonts}
\usepackage{algorithmic}
\usepackage{graphicx}
\usepackage{textcomp}
\usepackage{xcolor}
\usepackage{subcaption}

\def\BibTeX{{\rm B\kern-.05em{\sc i\kern-.025em b}\kern-.08em
    T\kern-.1667em\lower.7ex\hbox{E}\kern-.125emX}}

\makeatletter
\newcommand{\linebreakand}{%
  \end{@IEEEauthorhalign}
  \hfill\mbox{}\par
  \mbox{}\hfill\begin{@IEEEauthorhalign}
}
\makeatother

\begin{document}

\title{Research on Large Language Model Cross-Cloud Privacy Protection and Collaborative Training based on Federated Learning\\}

\author{

\small 

\begin{tabular}[t]{c@{\extracolsep{8em}}c} 

1\textsuperscript{st} Ze Yang \textsuperscript{*}  & 2\textsuperscript{nd} Yihong Jin \\
\textit{University of Illinois Urbana-Champaign} & \textit{University of Illinois Urbana-Champaign} \\
Champaign, IL 61801, USA & Champaign, IL 61801, USA \\
\textsuperscript{*}Corresponding Author: zeyang2@illinois.edu & yihongj3@illinois.edu\\

\\

3\textsuperscript{rd} Yihan Zhang & 4\textsuperscript{th} Juntian Liu \\
\textit{University of Illinois Urbana-Champaign} & \textit{University of Illinois Urbana-Champaign} \\
Champaign, IL 61801, USA & Champaign, IL 61801, USA \\
yihanz8@illinois.edu & jl203@illinois.edu  \\

\\

5\textsuperscript{th} Xinhe Xu\\  
\textit{University of Illinois Urbana-Champaign}\\ 
Champaign, IL 61801, USA \\ 
xinhexu2@illinois.edu \\ 

\end{tabular}
}

\maketitle

\begin{abstract}
The fast development of large language models (LLMs) and popularization of cloud computing have led to increasing concerns on privacy safeguarding and data security of cross-cloud model deployment and training as the key challenges. We present a new framework for addressing these issues along with enabling privacy preserving collaboration on training between distributed clouds based on federated learning. Our mechanism encompasses cutting-edge cryptographic primitives, dynamic model aggregation techniques, and cross-cloud data harmonization solutions to enhance security, efficiency, and scalability to the traditional federated learning paradigm. Furthermore, we proposed a hybrid aggregation scheme to mitigate the threat of Data Leakage and to optimize the aggregation of model updates, thus achieving substantial enhancement on the model effectiveness and stability. Experimental results demonstrate that the training efficiency, privacy protection, and model accuracy of the proposed model compare favorably to those of the traditional federated\enspace learning method.
\end{abstract}


\section{Introduction}
At present, with the rapid advancement in AI technology, large language models (LLMs) have made significant progress in various fields including natural language processing, machine translation, and dialogue systems. These models are essentially trained both semantic, inference, and generation capabilities and they have outperformed traditional methods in various complex tasks with multiple data points  \cite{zhu2023investigation, deng2024composerx, ding2024enhance, DBLP:conf/kdd/ZhengJLTH24,  DBLP:journals/corr/abs-2412-08174}. Such as, the most common used models in information extraction, sentiment analysis, text generation, and other tasks \cite{DBLP:journals/corr/abs-2410-12126, li2025language}; effectively promote the intelligent process of various industries. 

However, the training demands of these mammoth language models become even more grand as the model size grows. Because of the enormous amount of parameters, training these models requires very rich computing resources and a lot of labeled data. For example, data support of cross-domain models must be provided on a global scale, and it would be impossible for a single computing center or data storage platform to provide data needed for cross-domain models \cite{yang2024research}.

Therefore, cross-cloud platform collaborative training has become an increasingly important solution. By performing training tasks across multiple cloud service providers, the model fully utilizes distributed computing resources, lowers training costs, and enhances computing efficiency \cite{putri2025multi, DBLP:journals/corr/abs-2410-17576}. Some cloud platforms might have hardware advantages, while others might have closer-in data sources or services. Users can quintuple data and computing resources by working across clouds. 

However, besides the technical challenges, cross-cloud collaborative training also raises several critical issues like privacy preservation and data security. The rules for storing and processing data on different platforms are completely different in a cross-cloud environment. How to ensure efficient and seamless cooperation between different clouds and ensure that users' private data is not leaked has become an urgent problem to be solved \cite{liu2024recent}. For some fields with stringent data privacy protection, such as healthcare \cite{ji2024mitigating, du2024embracing, yi2018enhance, parker2024mapping}, finance, etc. that cannot share data, the privacy protection mechanism of cross-cloud joint training solution is of great significance.

In common large language model training methods, the centralized approach is adopted, that is, a machine data center or cloud platform (must achieve centralized data storage and processing) keeps all data, and directly accesses and processes all data, so that the undecentralized training has a performance guarantee. But this consolidating of data management is also a serious privacy issue and a concern for data sovereignty. 

User data can become compromised by third parties, and in fact, due to data access, a malicious attacker may be able to get sensitive information. With the continuous strengthening of data privacy supervision in various countries, the models' training without violating the user's privacy has become an important proposition that AI technology must confront in practical applications \cite{alhosban2024cvl, protocf, preprec, infomotif, DBLP:journals/corr/abs-2412-17336, he2024givestructuredreasoningknowledge,lu2023deep}. To overcome this problem, federated learning, a novel distributed machine learning approach, has gradually emerged as an effective approach to address the concerns of privacy protection and data sharing.

In federated learning, the centralized server sends the up-to-date models of the previous round to each client for training, but there is no need to send the original data to the central server, and only the model lmple updates (gradients) can be sent to the server for aggregation, which is mainly to protect the privacy of the data. So sensitive data is kept local and never actually leaks to servers or any other cloud platforms. While federated learning has a robust guarantee for privacy protection, it still encounters numerous challenges in application scenarios of cross-cloud platform collaborative training. 

\section{RELATED WORK}
Above all, reacting to the shift in resource allocation and policy priorities in public health, Behari et al. \cite{behari2024decision} applied Decision Language Model (DLM) to dynamic uncertain RMAB task with new framework. This approach leverages LLMs as automated planners and processes human language commands to generate reward functions that promote the adjustment of public health policy in a multi-agent reinforcement learning setting. Xu et al. \cite{xu2024sparsebf} proposed SparseBF, an improved Bloom Filter (BF) encoding method for privacy-preserving record linkage (PPRL) on sparse data. SparseBF achieves these subgoals via three contributions including a hybrid storage scheme that can flexibly choose a more efficient storage component, an adaptive compression method based on spatial compression for sparse row storage, and SIMD vector instruction support for accelerating computation process.

Chen et al. \cite{chen2024privacy} proposed a Privacy Passport, a formal privacy protection mechanism for the inter-domain data sharing. Advances in big data and cross-domain collaboration have greatly promoted the development of traditional privacy-preserving methods, where challenges such as model parameter sharing, privacy constraints and communication costs arise. A data tokenization-based solution that provides a secure and scalable way to share data across domains and institutions while preserving privacy. 

Additionally, Xiang et al. \cite{xiang2022nebula} proposed the Nebula-I framework for collaboratively training deep learning models in a low-bandwidth cloud cluster environment. As collaboration across cloud platforms becomes more and more important, existing cloud cluster training methods face significant challenges with limited bandwidth. Nebula-I overcomes the bottleneck of data transmission in a low-bandwidth environment and optimizes communication costs and computing efficiency by implementing efficient model training between multi-platform cloud clusters. Furthermore, Jinhong \cite{jinhong2024cross} develops an information security strategy for cross-platform multi-terminal collaboration software, aiming to ensure the integrity, confidentiality, and availability of in large-scale network environment data. The popularization of cloud computing and cross-platform software, how to between platform terminals or the cloud sharing sensitive data, at the same time to ensure data security has become key problem of information security field. 

\section{METHODOLOGIES}
\subsection{Cryptographic gradient update mechanism}
In traditional federated learning frameworks, clients upload model gradients or parameters directly, which can lead to the risk of data leakage. In order to ensure data privacy, we use Homomorphic Encryption technology to protect the gradient data of the client through encryption operations. Specifically, assuming that the gradient of client $k$ at time $t$ is $\nabla L_k(w_k)$, and the cryptographic operation is implemented by the homomorphic cryptographic function $\mathcal{E}(\cdot)$

\begin{equation}
\mathcal{E}(\nabla L_k (w_k))\quad\text{for each client } k \in \{1,2,\dots,K\}.
\label{eq:1}
\end{equation}

In order to improve the computational efficiency, we divide the cryptographic gradient into multiple small blocks for parallel calculation, and the cryptographic gradient of each block is Equation 2:

\begin{equation}
\mathcal{E}(\nabla L_k (w_k)) = \bigoplus_{i=1}^{n_k} \mathcal{E}(\nabla L_{k,i} (w_{k,i})),
\label{eq:2}
\end{equation}

where $n_k$ is the number of gradient blocks for client $k$, and $\nabla L_{k,i} (w_{k,i})$ is the gradient for the $i$-th block of the $k$-th client. In this way, distributed encryption can reduce the computational burden of encryption for each client and improve data transmission efficiency. When aggregating encrypted gradients on the server side, use the following encryption aggregation formula to merge the encrypted gradients of each client, as shown in Equation 3:

\begin{equation}
\mathcal{E}(\nabla L_t (w_t)) = \frac{1}{K} \sum_{k=1}^{K} \mathcal{E}(\nabla L_k (w_k)).
\label{eq:3}
\end{equation}

This formula realizes the weighted average of the cryptographic gradients uploaded by the client, so as to generate the cryptographic gradients of the global model to ensure that the original data is not leaked during the whole training process.
In order to further improve the accuracy and robustness of model training, we designed a hybrid aggregation method that combines encrypted gradients and conventional gradients. At each aggregation, we weighted the model updates and cryptographic gradients for each client according to a dynamic weighting mechanism. Specifically, the updated formula for the global model is Equation 4:

\begin{equation}
w_t = \frac{\sum_{k=1}^{K} \left( \omega_k \cdot \mathcal{E}(w_k (t)) + (1 - \omega_k) \cdot w_k (t) \right)}
{\sum_{k=1}^{K} \left( \omega_k + (1 - \omega_k) \right)}.
\label{eq:4}
\end{equation}

where $w_k$ is the weighting coefficient of client $k$, which dynamically adjusts based on factors such as the client's local loss, data size, and network bandwidth. In this formula, $\mathcal{E}(w_k (t))$ denotes the encrypted model update, and $w_k (t)$ is the unencrypted original model update. By using a hybrid weighting approach, we can further optimize the training effect of the global model while ensuring privacy protection.

To improve the adaptability of cross-cloud synchronization in complex network environments, we plan to introduce iperf, a real-time network monitoring tool to accurately measure key parameters such as network latency, bandwidth, and jitter. Based on this real-time data, we can design a dynamic adjustment algorithm to automatically optimize synchronization strategy.

\subsection{Dynamically weighted aggregation mechanism}
In order to further optimize the effect of cross-cloud platform collaborative training, we introduce a dynamic weighted aggregation mechanism to cope with the differences in data heterogeneity, network bandwidth constraints, and client computing power. In traditional federated learning, gradient or model updates for each client are aggregated with the same weight, which ignores the contribution of differentiation between clients. Our dynamically weighted aggregation mechanism adjusts the weights for each client based on several factors: the training loss of the local model $L_k (w_k)$, the data size of the client $|D_k|$, and the bandwidth of the client $B_k$. The weight $\omega_k$ of the client $k$ is defined as Equation 5:

\begin{equation}
\omega_k = \frac{L_k (w_k)}{|D_k| \cdot B_k^{\alpha}},
\label{eq:5}
\end{equation}

where $\alpha$ controls the effect of bandwidth on the weights. The implication of this formula is that clients with less training loss, larger datasets, and wider bandwidth will be given higher weight, making their updates contribute more to the global model. With this weighting, we are able to control the role of each client in the training process more precisely. The updated formula for the global model is Equation 6: 

\begin{equation}
w_t = \frac{\sum_{k=1}^{K} \omega_k w_k (t)}{\sum_{k=1}^{K} \omega_k}.
\label{eq:6}
\end{equation}

This formula aggregates the model updates of each client through dynamic weighting so that the training resources can be reasonably allocated and the accuracy of the model can be optimized during global model training.

Collaborative training across cloud platforms faces problems such as bandwidth constraints and network latency, so latency optimization and bandwidth management are required for data synchronization. We design a synchronization mechanism based on real-time network state, considering the synchronization delay $T_{\text{sync}}$ and bandwidth $B_{\text{could}}$ of each cloud platform, and calculate the total delay of cross-cloud synchronization by the following equation 7:

\begin{equation}
T_{\text{total}} = \sum_{i=1}^{N} \left( T_{\text{sync},i} + \frac{T_{\text{could}}}{B_{\text{could}}} \right).
\label{eq:7}
\end{equation}

where $N$ is the number of cloud platforms involved in cross-cloud training, $T_{\text{sync},i}$ is the synchronization delay of the $i$-th cloud platform, $T_{\text{could}}$ is the training time of each platform, and $B_{\text{could}}$ is the bandwidth limit of each platform. In order to reduce the synchronization delay, we introduce an asynchronous synchronization mechanism with weights and introduce dynamic weights $\omega_{\text{sync},i}$ for the synchronization requirements of each cloud platform. The weighted formula for asynchronous synchronization is Equation 8:

\begin{equation}
T_{\text{sync,weighted}} = \sum_{i=1}^{T} \omega_{\text{sync},i} T_{\text{sync},i}.
\label{eq:8}
\end{equation}

By dynamically adjusting the weights, we can optimize the synchronization process according to the bandwidth, latency, and load conditions of different cloud platforms, reducing the overhead of data synchronization.

\section{EXPERIMENTS}
\subsection{Experimental setup}

We utilized the MIMIC-III dataset to validate the proposed experimental dataset using the federated learning framework. It includes multimodal, high heterogeneity, privacy-sensitive, and highly imbalanced data with a large amount of clinical data from patients in the intensive care unit, which consists of diverse physiological monitoring, laboratory results, and treatment details, as the clinical data is from a multimodal system. The data was standardized, missing values were completed and divided, a federated learning cloud platform simulation.
To thoroughly assess the effectiveness of the proposed innovative framework that leverages federated learning, we chose four comparative methods related to it: centralized training method, traditional federated learning (FL), homomorphic encryption-based federated learning (HE-FL), and differentially private federated learning (DP-FL). 

\subsection{Experimental analysis}
Following Figure \ref{fig:data lake rate} demonstrates the intuitive comparison of the privacy protection capability of each method, as well as along with the increased number of training rounds, the leakage rate of all methods should gradually decline, but the decline rate of different methods is different, and it can also reflect the specific privacy protection effect and training cost. The practical privacy saving will give a better perspective for comparing the benefits of Ours with other methods in terms of actual privacy protection by looking at how they perform when the number of training rounds increases in terms of data breaches.

\begin{figure}[h!]
  \centering
  \begin{subfigure}[T]{1\linewidth}
    \includegraphics[width=\linewidth, height=0.8\linewidth]{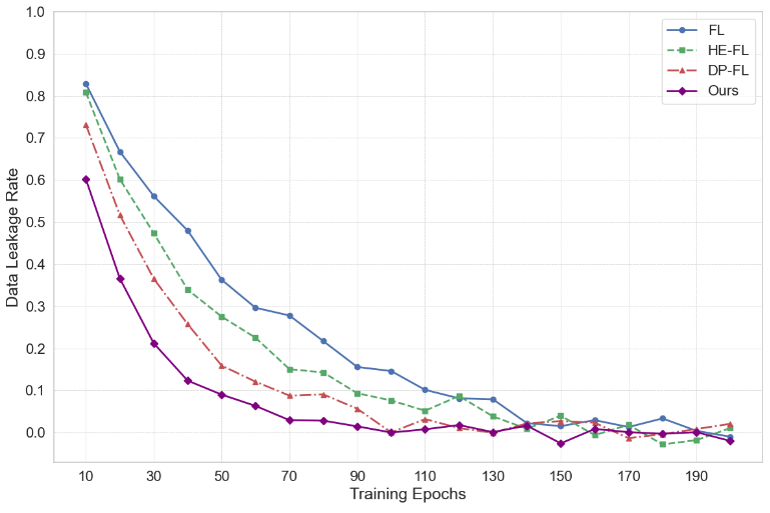}
  \end{subfigure}
  \caption{Comparison of Data Leakage Rate}
  \label{fig:data lake rate}
\end{figure}

\begin{figure}[h!]
  \centering
  \begin{subfigure}[T]{1\linewidth}
    \includegraphics[width=\linewidth, height=0.8\linewidth]{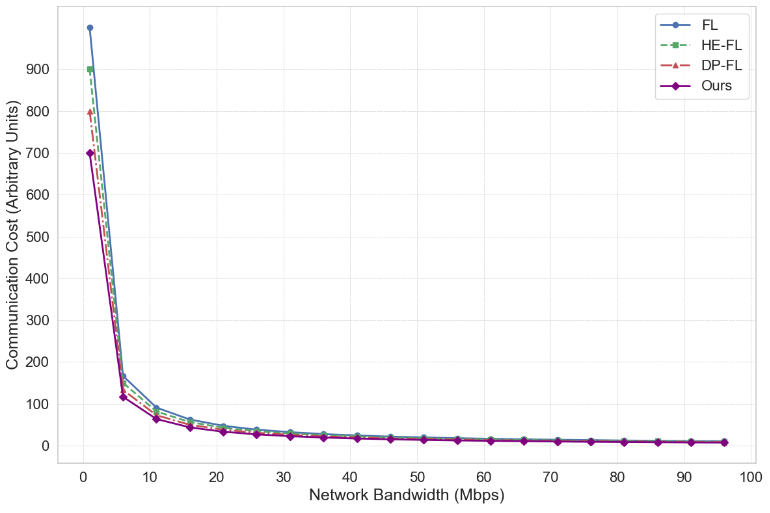}
  \end{subfigure}
  \caption{Comparison of Communication Cost}
  \label{fig:Communication Cost}
\end{figure}

The communication cost is mainly to measure the resource consumption different methods in the cross-cloud collaborative training process, especially under various network bandwidth conditions the data transmission overhead. The communication cost is much less in the communication cost comparison shown in Figure \ref{fig:Communication Cost}, which indicates that Ours method is more effective than other comparison methods. While all methods see a lower communication cost with increasing bandwidth, the Ours method benefits the most, demonstrating that it can exploit its reduced redundant communication and optimised data transfers.

\begin{figure}[h!]
  \centering
  \begin{subfigure}[T]{1\linewidth}
    \includegraphics[width=\linewidth, height=0.8\linewidth]{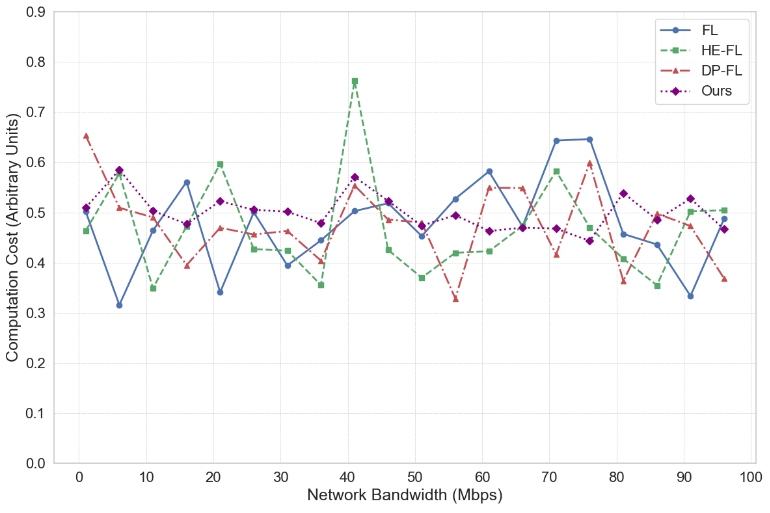}
  \end{subfigure}
  \caption{Comparison of Computation Cost}
  \label{fig:Computation Cost}
\end{figure}

We evaluate the computational costs of four methods under different network bandwidth conditions. The results of Figure \ref{fig:Computation Cost} show that the computational cost of each method decreases with the increase of bandwidth. However, the computational cost of the Ours method is consistently lower than that of the other three methods, and it shows better performance. This shows that the Ours method can reduce the computing cost more effectively and has a higher cost performance while increasing the bandwidth. 

In small-scale datasets, the speed-up(reverse time) of these block encryption and parallel computing optimization methods is approximately $20\%$ compared with conventional AES encryption method. Homomorphic encryption, in particular, is nearly 2 times of computation time to encrypt and reverse encryption of homomorphic encryption than AES encryption. Block encryption and parallel computing has begun to show its apparent advantage with the increasing amount of dataset. The computational efficiency of traditional AES encryption is improved by more than $30\%$, especially in the decryption process, on a data set of 1 million records, where block encryption and optimization of parallel computing were applied. On the 10 million-scale dataset, our optimization strategy with the combination of block encryption and parallel computing is greatly scalable, and the computation time reduces about $45\%$ comparing to the traditional AES encryption.

\subsection{CONCLUSION}
In conclusion, They utilize the federated learning, a new framework, to provide the training between different cloud while still ensuring the confidentiality of private data. Thus, open-source technologies use a combination of advanced cryptography, dynamic model aggregation, an enhanced class of generalized stochastic modeling and cross cloud data synchronization to add security, efficiency and scalability. Our method shows superior performance on experimental results than the traditional one on the aspect of communication cost. Future work can consider sacrifice of the learning performance, especially in limited bandwidth.

\renewcommand{\bibfont}{\footnotesize}

\footnotesize{
\bibliographystyle{IEEEtran}
\bibliography{main}
}

\end{document}